\documentclass[
reprint,
amsmath,
amssymb,
aps,
prl,
superscriptaddress,
longbibliography,
floatfix,
nobalancelastpage,
]{revtex4-2}

\usepackage{graphicx}% Include figure files
\usepackage{dcolumn}% Align table columns on decimal point
\usepackage{bm}% bold math
\usepackage{epstopdf}
\usepackage{float}
\usepackage{braket}
\usepackage{dsfont}
\usepackage{amsmath}
\usepackage{siunitx} % typesets numbers with units very nicely
\usepackage[italicdiff]{physics}
\usepackage[T1]{fontenc}
\usepackage[dvipsnames]{xcolor}
\usepackage{color,soul}
\usepackage{lipsum}
\usepackage[colorlinks=true, allcolors=blue]{hyperref}

\graphicspath{{figures/}}

\begin{document}

\title{Measuring and correcting nanosecond pulse distortions in quantum-dot spin qubits}

\author{Jiheng Duan}
\thanks{These authors contributed equally.}

\author{Fernando Torres-Leal}
\thanks{These authors contributed equally.}

\author{John M. Nichol}
\email{john.nichol@rochester.edu}
\affiliation{Department of Physics and Astronomy, University of Rochester, Rochester, NY, 14627 USA}
\affiliation{University of Rochester Center for Coherence and Quantum Science, Rochester, NY, 14627, USA}

\begin{abstract}
Gate-defined semiconductor quantum dots utilize fast electrical control to manipulate spin and charge states of individual electrons.
Electrical pulse distortions can limit control fidelities but are difficult to measure at the device level.
Here, we use detuning-axis pulsed spectroscopy to characterize baseband pulse distortions in a silicon double quantum-dot. 
We extract the gate-voltage impulse response and apply a digital pre-distortion filter to eliminate pulse distortions on timescales longer than 1~ns.
With the pre-distortion, we reduce the frequency chirp of coherent exchange oscillations in a singlet-triplet qubit. 
Our results suggest a scalable and tuning-efficient method for characterizing pulse distortions in quantum-dot spin qubits.
\end{abstract}


\maketitle

\emph{Introduction--}
Semiconductor spin qubits are a promising platform for high fidelity quantum information processing~\cite{burkard2023semiconductor}. 
Most semiconductor spin qubits rely on fast electrical control through various spin-charge coupling mechanisms. 
For example, exchange coupling is most commonly generated through baseband voltage pulses applied to barrier gates~\cite{reed2016reduced,martins2016noise}. 
Baseband voltage pulses are also used for single-qubit operations for two- and three-spin qubits~\cite{petta2005coherent,divincenzo2000universal}, and ac voltage pulses drive single-spin rotations~\cite{pioro2008electrically}. 

However, such electrical control signals inevitably suffer from pulse distortions, which originate from the various elements and properties of wiring between room temperature electronics and cryogenic quantum devices, including bias-tees, wirebonds, and device packaging. 
These distortions, which include frequency-dependent attenuation~\cite{wigington2007transient} and reflections~\cite{pozar2021microwave}, are especially severe for baseband pulses and result in pulse-dependent coherent errors during qubit operations. Often, these errors cannot be distinguished from other coherent errors through randomized benchmarking or tomography protocols~\cite{gustavsson2013improving, sheldon2016characterizing, tanttu2024assessment}. 
Moreover, high-fidelity control methods, including numerically optimized pulses~\cite{khaneja2005optimal, caneva2011chopped, calderon2019fast, yang2017achieving, hansen2021pulse, rimbach2023simple}, dynamically corrected gates~\cite{wang2012composite, kestner2013noise, yang2019silicon, walelign2024dynamically},  and leakage-suppressing pulses~\cite{rimbach2023simple, polat2025pulse, motzoi2009simple}, often depend critically on precise pulse shaping and are thus highly susceptible to pulse distortions.

While distortions ``above'' the device level can be calibrated at room temperature~\cite{cerfontaine2020closed}, such calibrations often do not include the quantum device itself and do not include any changes that occur in cooling to cryogenic temperatures. 
An ideal approach is to create a ``cryoscope''~\cite{rol2020time}, which can measure pulse distortions in-situ. 
Ramsey-interferometry-based cryoscopes have been well developed in superconducting flux qubits~\cite{gustavsson2013improving}, tunable transmons~\cite{rol2019fast, rol2020time, li2019realisation, guo2024correction, guo2024universal, hyyppa2024reducing, sung2021realization}, and tunable couplers~\cite{sung2021realization, chen2025efficient}. 
Methods in semiconductor spin qubits typically employ exchange-based qubit operations ~\cite{foletti2009universal, barnes2020correcting, philips2022universal, ni2025correcting, mkadzik2025operating} to resolve how voltage pulses depend on time.

Here, we introduce a cryoscope realized in a semiconductor double quantum-dot (DQD) that does not involve any coherent measurements. Instead, our cryoscope measures fast, incoherent charge dynamics through detuning-axis pulsed spectroscopy~\cite{chen2021detuning}.
This approach resolves distortions on individual plunger gates on the device level at base temperature.
A practical advantage of using incoherent charge dynamics rather than coherent spin dynamics is that the calibration can be performed at an early stage of device characterization, before a spin qubit has been tuned up
We characterize all drive-line impulse response functions and develop digital pre-distortion filters to correct them. 
Using this approach, we correct pulse distortions down to one nanosecond, limited by our hardware sampling rate, and we demonstrate a reduction in the frequency chirp of exchange oscillations in a singlet-triplet qubit. 
In total, our results demonstrate a scalable and tuning-efficient method to characterize and correct pulse distortions for semiconductor spin qubits.

\begin{figure}[ht]
{\includegraphics{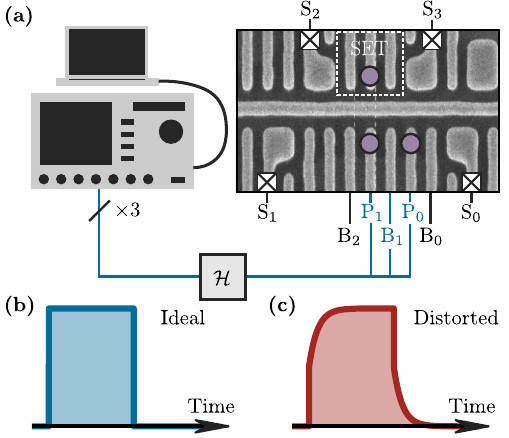}} 
\caption{
(a) Experimental setup. 
A scanning electron microscope image of a device similar to the one used here.
Ohmic contacts are indicated with white boxes. 
The DQD is confined in dots 0 and 1, with plunger gates  P$_0$ and P$_1$, respectively, and B$_1$ is the inter-dot tunneling gate. 
$S_0$, ..., $S_3$ are ohmic contacts. 
We use a single-electron transistor (SET) for spin and charge readout.
P$_0$, P$_1$, and B$_1$ are connected through individual drive lines to room temperature electronics. 
Each drive line has its own transfer function $\mathcal{H}$, describing the distortions of that line.  
(b) The ideal square pulse, programmed inside the AWG.
(c) Example distorted square pulse, delivered to the DQD.
}
\label{fig:setup}
\end{figure}

% \section{Characterizing distortions}
\emph{Characterizing distortions--}
We characterize pulse distortions in a Intel Tunnel Fall triple-dot device fabricated on a Si/SiGe heterostructure~\cite{neyens2024probing, george202412}, as shown in Fig.~\ref{fig:setup}(a). We use radio-frequency (rf) reflectometry~\cite{connors2020rapid} for spin- and charge-state readout. Voltage pulses on plunger gates P$_0$ and P$_1$ and barrier gate B$_1$ are driven by a room-temperature arbitrary waveform generator (AWG) and travel through various coaxial cables, wirebonds, and gate structures before reaching the quantum dots. We assume that for each gate, the signal reaching the dots is characterized by an impulse response function $h(t)$, whose Fourier-transform is the transfer function $\mathcal{H}$ [Figs.~\ref{fig:setup}(b)-(c)].

\begin{figure*}[!ht]
{\includegraphics{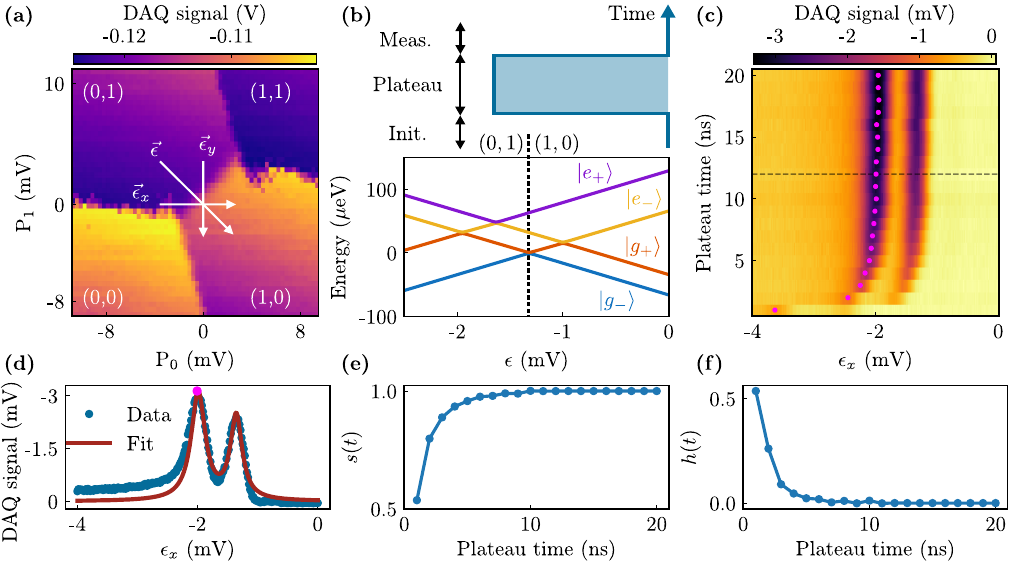}}
\caption{
(a) Charge stability diagram. 
The detuning axis is defined along its unit vector $\hat{\epsilon} = (0.74788, -0.66384)$, where $\vec{\epsilon}_x$ and $\vec{\epsilon}_y$ are the horizontal and vertical components, respectively.
We define $\epsilon=0$ at the initialization and measurement point, where the three vectors intersect. 
(b) Energy level diagram of the DQD along the detuning axis. 
The dashed line in the middle indicates the charge transition boundary.
The DAPS pulse sequence, consisting of initialization, evolution, and measurement segments, is illustrated along the time axis on the right. 
(c) One-electron DAPS data along the horizontal $\vec{\epsilon}_x$ axis. 
The extracted peaks of the $|g_-\rangle\leftrightarrow|e_+\rangle$ transitions are highlighted. 
For each horizontal line, we subtract the value at  $\epsilon_x = 0$~mV.
(d) Horizontal trace of the DAPS data from (c) along the dashed line. 
(e) Extracted step response function $s(t)$ and (f) impulse response $h(t)$ of the P$_0$ drive line.
}
\label{fig:DAPS}
\end{figure*}

We use detuning-axis pulsed spectroscopy (DAPS)~\cite{chen2021detuning} to characterize the step-response function $s(t)=\int h(t) dt$ of each drive line. To implement DAPS in our device, we tune the DQD to the charge-transition boundary between the $(1,0)$ and $(0,1)$ charge configurations. After initializing the DQD in the $(1,0)$ configuration, we pulse the detuning $\epsilon(t)$ for a time $t$ with varying amplitudes and record the charge transition probability. Figure~\ref{fig:DAPS}(b) shows the relevant DQD energy levels and the DAPS pulse sequence.
When the detuning pulse reaches one of the avoided crossings between the $(1,0)$ and $(0,1)$ charge states, a measurement of the charge state after the pulse will have a non-zero probability to yield $(0,1)$. 
In our experiment, changes in the down-converted reflectometry signal reflect changes in the charge state of the DQD.  

The central idea of this work is that by finding $\epsilon^{max}(t)$, the pulse amplitude giving the largest $(0,1)$ measurement probability (the DAPS signal) as a function of time $t$, we can reconstruct the step response of our experimental wiring, up to and including the DQD confinement gates. 
To understand our approach, consider a simplified scenario where we apply an ideal step-function detuning pulse $\epsilon\theta(t)$ with amplitude $\epsilon$, where $\theta(t)$ is the Heaviside step function. 
This ideal step is distorted by the drive line to $\epsilon(t)=\epsilon s(t)$, where $s(t)$ is the step response. 
Considering only two levels [e.g., $\ket{g-}$ and $\ket{e-}$ in Fig. ~\ref{fig:DAPS}(b)] with detuning $\epsilon(t)$, at each instant in time, we can model the relaxation rate of the system as
\begin{equation}\label{eqS:analytical_gamma}
    \Gamma(t) = \frac{t_c^2\kappa}{\kappa^2 + \left[\epsilon(t) - \epsilon_0\right]^2},
\end{equation}
where these two levels are on resonance at $\epsilon_0$.
This expression assumes that the charge dephasing rate $\kappa$ is much larger than the $e \to g$ relaxation rate and the inter-dot tunnel coupling $t_c$. Let us suppose that the system is initially prepared in the ground state, and let us denote the probability that an energy measurement of the system yields the eigenvalue corresponding with $\ket{g} (\ket{e})$ as $p_g(p_e)$. Let us define the DAPS signal $w=2p_e$, which in practice ranges between 0 and 1. As described in the Supplemental Material~\cite{supplement}, $w$ obeys $\dot{w}=-\Gamma(t)(w-1)$.%
\nocite{tikhonov1963solution, hansen2010discrete, vogel2002computational, bylander2011noise, connors2019low, ye2024characterization, yoneda2023noise}% Assuming that $\dot{\epsilon}(t)>0$ is a monotonically decreasing function of time (e.g., the distortion is smooth and has no oscillations), and considering only charge-noise induced state changes, we find
\begin{equation}\label{eq:pop_solution}
    w(t,\epsilon) = 1-\exp\left[{-\int_0^{\epsilon s(t)} \frac{\Gamma(\varepsilon)}{\dot{\epsilon}(s^{-1}(\varepsilon/\epsilon))} d\varepsilon}\right].
\end{equation}
In the limit of small $\kappa$ and $t_c$, we may approximate $\Gamma(\epsilon)\approx \pi t_c^2 \delta(\epsilon-\epsilon_0)$, in which case 
\begin{equation}\label{eq:population_relation}
    w(t,\epsilon) \approx 
    \begin{cases}
        1-\exp \left( -\frac{\pi t_c^2}{\dot{\epsilon}(t')}\right) & \textnormal{if} \qquad t' \leq t \\
        0  & \textnormal{otherwise}
    \end{cases},
\end{equation}
where $t'=s^{-1}(\epsilon_0/\epsilon)$ is the time at which the distorted detuning pulse crosses $\epsilon_0$. 

The physical content of Eq.~(\ref{eq:population_relation}) is as follows.
For a fixed pulse duration $t$, sweeping the pulse amplitude $\epsilon$ changes the time at which the distorted pulse $\epsilon s(t)$ crosses the resonance $\epsilon_0$. 
Small-amplitude pulses never reach resonance and produce no signal.
Large-amplitude pulses sweep through resonance early and quickly, spending little time there, and giving little signal.
The amplitude $\epsilon^\text{max}(t)$ that maximizes the signal is the one for which the distorted pulse just reaches $\epsilon_0$ at the end of the plateau, i.e., where $\dot{\epsilon}$ at the crossing is the smallest and the system dwells longest at resonance. 
This condition is $s(t)=\epsilon_0/\epsilon^\text{max}(t)$, so tracking $\epsilon^\text{max}(t)$ directly reconstructs the step response~(see Supplemental Material).

To implement this scheme in practice for each of the plunger gates, we define the detuning direction $\vec{\epsilon}$ perpendicular to the $(1,0)$-$(0,1)$ tunneling transition. 
The pulse sequence starts with a 30~$\mu$s initialization period at $\epsilon=0$, after which the system empirically has a large probability to be measured in the ground valley state $|g_-\rangle$. (The idling tuning of the device features a large coupling between dot 0 and its reservoir to facilitate rapid initialization.) 
Figure~\ref{fig:DAPS}(c) shows the result of DAPS measurements along the $\vec{\epsilon}_x$ direction, which feature a pulse applied to the P$_0$ gate only. 
The two peaks indicate the level-crossing between $|g_-\rangle\leftrightarrow|e_-\rangle$ and $|g_-\rangle\leftrightarrow|e_+\rangle$, respectively, and the splitting between the two peaks indicates the valley splitting in dot 1~\cite{chen2021detuning}. 
By fitting the data with a double-Lorentzian model [Fig.~\ref{fig:DAPS}(d)], we extract $\epsilon^{max}(t)$, the pulse amplitudes giving the largest signal.
We assume that $\epsilon_0=\epsilon^{max}(20~\textnormal{ns})$ and use the formalism discussed above to extract $s(t)$ and $h(t)=ds/dt
$. 
Both responses are shown in Figs.~\ref{fig:DAPS}(e)-(f).
We apply the same method to the P$_1$ gate. To characterize the distortion of B$_1$, we tune up a double dot under B$_1$ and B$_2$ and repeat (see Supplemental Material). 

We note that in practice, the extracted impulse response is not exact, because the applied control pulses have finite rise and fall times due to hardware bandwidth limitations and distortions. 
Because of the finite rise and fall times, the measured DAPS signal reflects a time-integrated response of the instantaneous relaxation rate rather than the instantaneous rate itself. Equivalently, in real systems, the approximation $\Gamma(\epsilon)\approx \pi t_c^2 \delta(\epsilon-\epsilon_0)$ never holds exactly.  
As a result, the pulse amplitude that maximizes the integrated population change does not coincide exactly with the amplitude that first brings the system to the avoided crossing, and the step response extracted from $\epsilon_\text{max}(t)$ will have systematic errors. 
As discussed in the Supplemental Material (Fig.~S4), this results in a pre-distortion filter that typically introduces a slight overshoot in the actual waveforms. In the following, we manually adjust the pre-distortion filter to eliminate the overshoot (see Supplementary Figures~S3 and S5). This adjustment can also be performed automatically with filter regularization and a gradient-descent method (see the Supplemental Material).

We also note that the ability to precisely measure $\epsilon^{max}(t)$ depends both on the absolute signal level of the DAPS peak, as well as its width. 
In the model discussed above [Eq.~\eqref{eq:population_relation}], which applies when $\kappa \gg t_c$ and when both are small relative to experimental detuning changes, the signal level scales as $1-\exp \left( -\pi t_c^2/\dot{\epsilon}\right)$. 
Thus, on the one hand, the signal level increases with $t_c$. 
On the other hand, the width of the DAPS peak will also increase when the condition $\kappa \gg t_c$ is no longer satisfied, because contributions from the first-order asymptotic correction to the dephasing dynamics must be included (see Supplemental Material). 
Within this simplified picture, given a charge dephasing rate $\kappa$, the tunneling should be as large as possible while still satisfying $\kappa \gg t_c$. 
The resonance condition $\epsilon_0$ should then be set as large as possible to ensure a narrow peak width, but not so large that $t_c^2/\dot{\epsilon}$ reduces and the DAPS peak signal falls below the experimental noise level.  

\begin{figure}[t]
{\includegraphics{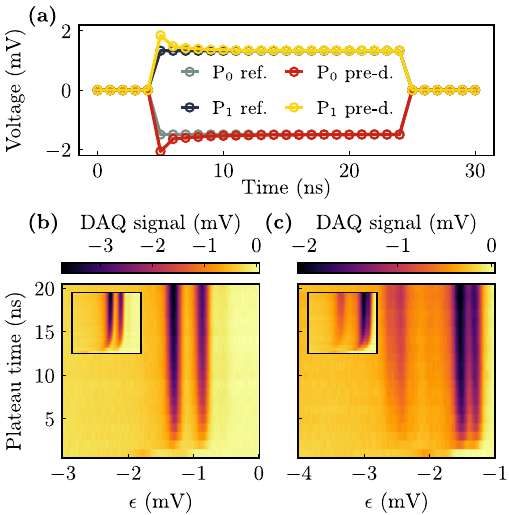}}
\caption{
(a) Square-pulse waveform with and without pre-distortion. 
(b) One-electron DAPS data along the detuning axis $\vec{\epsilon}$ with pre-distortion applied to both plunger gates. The inset shows the result without pre-distortion.
(c) Four-electron DAPS along the detuning axis $\vec{\epsilon}$ with pre-distortion. The inset shows the result without pre-distortion.
}
\label{fig:predistorted_DAPS}
\end{figure}

\emph{Correcting distortions--}
After measuring the impulse response of our drive lines, we correct the distortions using finite impulse response (FIR) filters to pre-distort digital waveforms at room temperature. 
We design these filters by writing the discrete impulse response convolution operator as a matrix and then inverting that matrix (see the Supplemental Material). 
As mentioned above, and as discussed in the Supplemental Material, the pre-distortion filter extracted from this method introduces an overshoot, which we correct by systematically adjusting the filter coefficients. (See the Supplemental Material for a description of this adjustment for the B$_1$ gate.)
In practice, we also apply this pre-distortion filter only to the first 20~ns of the square pulse where the distortions are predominant [Fig.~\ref{fig:predistorted_DAPS}(a)]. (This allows us to avoid recalibrating the DQD initialization and readout with the pre-distortion.)  

Figure~\ref{fig:predistorted_DAPS} shows the results of a DAPS experiment with pre-distortion corrections applied. 
We observe straight DAPS lines along the detuning axis $\vec{\epsilon}$, as shown in Fig.~\ref{fig:predistorted_DAPS}(b). 
The inset of Fig.~\ref{fig:predistorted_DAPS}(b) shows the result without pre-distortions. 
Note that these measurements were taken after calibrating and correcting the pre-distortion filters for both P$_0$ and P$_1$. 
The pulse distortions are largely eliminated down to about 1~ns, corresponding to the sampling rate of our AWG. 

Below, we discuss how the pre-distortion filter affects the performance of a $(1,3)$-$(0,4)$ singlet-triplet qubit~\cite{connors2022charge,harvey2017coherent}. (This four-electron singlet-triplet qubit features filled valley states in dot 0. Thus, the Pauli-spin-blockade readout is limited by the $s$-$p$ orbital energy spacing, instead of the valley splitting.) 
To verify that the pre-distortion does not depend on tuning or electron number, we tune the device to the $(1,3)$-$(0,4)$ charge transition and perform DAPS experiments using the same FIR filters calibrated in one-electron measurement. Figure~\ref{fig:predistorted_DAPS}(c) shows our results. 
As in the one-electron case, the pre-distortion appears effective at eliminating pulse distortions. 
The origin of the peak structure with four electrons is not entirely clear~\cite{chen2021detuning} but may result from the valley splitting of the dot 1, together with different possible valley configurations of three-electron singlet states or higher orbital energy spacings~\cite{liu2021magnetic,cai2023coherent}.

\emph{Mitigating ST$_0$ qubit frequency chirp--}
To verify that our pre-distortion calibrated from DAPS experiments can improve spin-qubit operations, we configure the four-electron DQD as a singlet-triplet qubit~\cite{connors2022charge} and measure pulsed exchange oscillations. 
With a small in-plane magnetic field (10 mT), we use adiabatic preparation and readout of Zeeman-gradient eigenstates, together with a combined detuning and barrier pulse to induce exchange~\cite{petta2005coherent,foletti2009universal}. 

\begin{figure}[t]
{\includegraphics{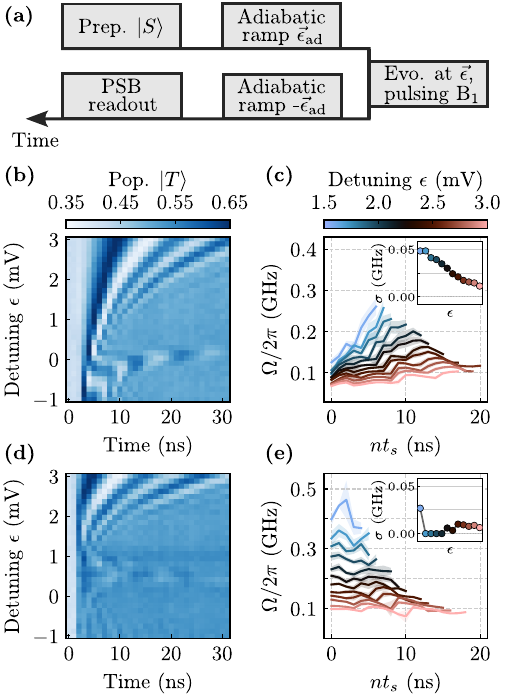}}
\caption{
(a) Sequence to measure pulsed exchange oscillations in a singlet-triplet qubit.
The adiabatic ramp position $\vec{\epsilon}_\text{ad}$ is along the detuning axis $\vec{\epsilon}$ with its amplitude $\epsilon_\text{ad}$.
(b), (d) Pulsed exchange oscillations (c) without, and (e) with pre-distortion. 
(c), (e) Oscillation frequencies for $\epsilon>1.5$~mV, extracted by fitting the data in (b) and (d) within sliding windows, respectively. 
Extracted data outside the coherence time at each detuning are omitted. 
The horizontal axes represent the initial timestamp $nt_s$ of the sliding windows with $t_s = 1$~ns and $n=0,1,2,\cdots$.
The error bars are the mean standard errors of the fitted frequencies.
The inset plots in (c), (e) show the calculated frequency deviation $\sigma$ without and with pre-distortion, respectively.
}
\label{fig:STqubit}
\end{figure}

We perform the sequence described in Fig.~\ref{fig:STqubit}(a) with a fixed barrier-gate pulse amplitude $75$~mV, and Figs.~\ref{fig:STqubit}(b) and (d) show the resulting exchange oscillations without and with pre-distortion, respectively. 
In both cases, we observe a peak-to-peak oscillation amplitude of about $0.3$. 
This relatively low amplitude is likely limited by the fidelity of the adiabatic preparation. 
The similar amplitude value between experiments indicates that the pre-distortion does not seriously limit the state preparation below the value set by the relatively small Zeeman gradient. 
A faint chevron pattern appears around $\epsilon = 0.5$~mV, likely a result of spin-orbit driven $|S\rangle\leftrightarrow|T_-\rangle$ oscillations.
Without the pre-distortion, the chevron pattern is ``curved,'' due to pulse distortions, as shown in Fig.~\ref{fig:STqubit}(b) (also see the Supplemental Material).

At larger values of the detuning, we observe exchange oscillations. 
Without the pre-distortion, we observe a pronounced frequency chirp in the exchange oscillations [Fig.~\ref{fig:STqubit}(b)]. 
To quantify the chirp, we fit the data at $\epsilon>1.5$~mV in a sliding single-period window to a sinusoidal function to extract the time-dependent frequency. 
Specifically, we fit the data in a window defined by $nt_s < t < T + nt_s$, where $T$ is approximately the oscillation period at each $\epsilon$, $t_s=1$ ns is the sampling time, and $n=0,1,2,\cdots$, to a single-frequency sinusoidal function $A \cos\left[\Omega (t-t_0)\right]+B$ with four free parameters. 
Figures~\ref{fig:STqubit}(c) and (e) show the extracted frequency $\Omega$ without and with the pre-distortion, respectively. 
We use the frequency deviation $\sigma=\sqrt{\text{Var}(\Omega)-(\delta\Omega)^2}/2\pi$ to quantitatively describe the chirp, where $\delta\Omega$ is the mean standard error of the fitted frequencies.
$\sigma\rightarrow0$ indicates less frequency chirp.
As illustrated in the inset plots of Figures~\ref{fig:STqubit}(c) and (e), the calculated $\sigma$ with pre-distortion is significantly smaller than the result without pre-distortion.  
The corrected detuning and tunneling pulses, compared with the distorted case, also result in a larger exchange coupling and shorter coherence time, as expected~\cite{reed2016reduced}.  
The resulting high frequency and short coherence time likely result in the spike at $\epsilon\approx1.5$~mV in Fig.~\ref{fig:STqubit}(e) due to failure of the period fitting.  
While our analysis method is not sensitive to sub-period chirp, the rapid chirp near $nt_s=0$ visible in Fig.~\ref{fig:STqubit}(c) also appears to diminish with pre-distortion in Fig.~\ref{fig:STqubit}(e). 
Overall, our analysis shows reduced frequency variations in time, suggesting that the pre-distortion reduces the chirp. 
Additional methods for analyzing the frequency chirp are discussed in Supplemental Material.

\emph{Conclusion--}
To summarize, we have developed a tuning-efficient cryoscope based on a semiconductor double quantum-dot that enables characterizing baseband pulse distortions at the device level. 
Using detuning-axis pulsed spectroscopy, we resolve the step response of individual gates in a quantum-dot device. 
Based on the measured response functions, we implement FIR filters to correct the pulse distortions.
In the future, it may be possible to design calibration waveforms or processes that minimize residual overshoots or are compatible with non-monotonic distortions such as ringing.
In the context of advancing spin-qubit-based quantum information processing, our results demonstrate a scalable approach for fast and accurate distortion characterization across all plunger or finger gates in large-scale quantum-dot processors.

\emph{Data Availability--}
The processed data that support the plots are available at Zenodo~\cite{duan_2026_18511000}. The raw data are available from the corresponding author upon request.

\emph{Acknowledgments--}
J. D. acknowledges discussions with Yichen Shi. Research was sponsored by the Army Research Office and was accomplished under Cooperative Agreement Number W911NF-22-2-0037 and Grant Number W911NF-23-1-0115. The views and conclusions contained in this document are those of the authors and should not be interpreted as representing the official policies, either expressed or implied, of the Army Research Office or the U.S. Government. The U.S. Government is authorized to reproduce and distribute reprints for Government purposes notwithstanding any copyright notation herein. We acknowledge support from Intel Corporation for providing the device.

\emph{Author Contributions--}
J.D. and F.T.L. formulated the experiment and carried out the measurements; J.D. and F.T.L. analyzed the data and contributed to the software infrastructure; J.D. wrote the paper and got feedback from all authors; J.M.N. supervised the research.

% Temporarily disable \addcontentsline (no \begingroup: a group would
% reset REVTeX's internal bibdata flag and duplicate \bibdata in main.aux)
% so that "References" stays out of the Supplementary Information contents list.
\makeatletter
\let\SM@addcontentsline\addcontentsline
\let\addcontentsline\@gobblethree
\bibliography{main}
\let\addcontentsline\SM@addcontentsline
\makeatother

% Supplementary information, appended to the end of the main text for arXiv.
\onecolumngrid
\newpage
\begin{center}
    \textbf{SUPPLEMENTARY INFORMATION}
\end{center}

\setcounter{figure}{0}
\setcounter{equation}{0}
\setcounter{table}{0}
\setcounter{section}{0}
\setcounter{secnumdepth}{3}  % prl class suppresses section numbers; restore them for the SM
\makeatletter
\renewcommand{\thesection}{S\@arabic\c@section}
\renewcommand{\thefigure}{S\@arabic\c@figure}
\renewcommand{\thetable}{S\Roman{table}}
\renewcommand{\theequation}{S\arabic{equation}}
\makeatother

\tableofcontents

% \clearpage
\section{Experiment setup}

\begin{table}[t]
    \centering
    \begin{tabular}{|c|c|c|}
        \hline
         Component & Manufacturer & Model \\
         \hline
         AWG & Tektronix & AWG5014C\\

         DC & DECADAC & - \\

         LO & Lab Brick & LSG-451\\

         ADC & Alazar Technologies & ATS9440\\

         RF switch & Mini-Circuits & ZASWA-2-50DR+\\

         Directional coupler & Mini-Circuits & ZEDC-15-2B\\

         Mixer & Mini-Circuits & ZX05-1LHW-S+\\

         Analog phase shifter & PULSAR & SO-06-411\\

         Room temperature amplifier & WENTEQ MICROWAVE & ABL0040-00-6010\\

         Low temperature amplifier & Cosmic Microwave Technology & CITLF SN980L\\
         \hline
    \end{tabular}
    \caption{Instrument information}
    \label{tab:S1_instrument}
\end{table}

\begin{figure}[ht]
\centering
{\includegraphics[width=0.5 \textwidth]{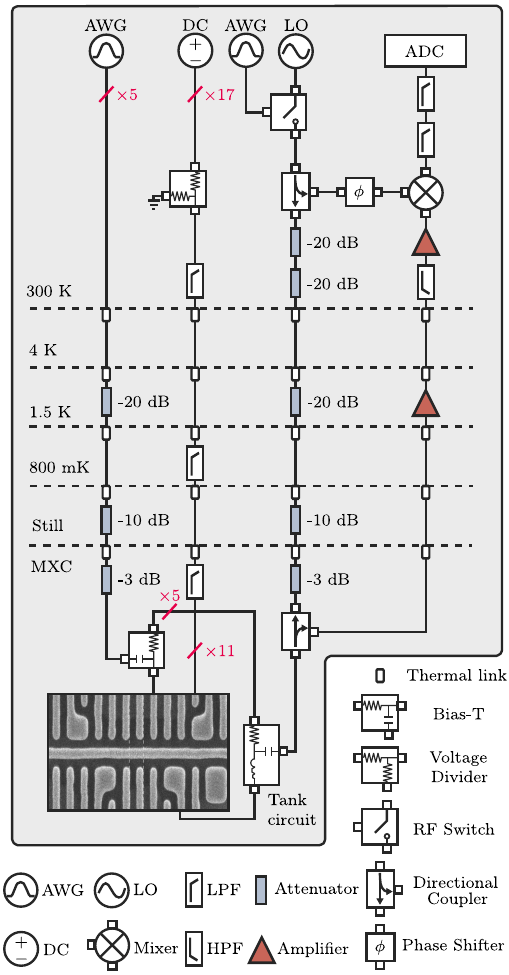}}
\caption{
Device and fridge wiring.
}
\label{figS:S1_fridge_wiring}
\end{figure}

We perform measurements on a Si/SiGe Intel Tunnel Fall triple-quantum-dot device~\cite{neyens2024probing,george202412} with 800-ppm isotopic purification, cooled in an Oxford Instruments Kelvinox MX400 dilution refrigerator.
The device wiring configuration and the naming scheme of the metallic gates are shown in Fig.~\ref{figS:S1_fridge_wiring} and Fig.~\ref{figS:S2_calibrate_B1_B2}(a), respectively.
DC voltages from room-temperature sources are attenuated by custom voltage dividers with a ratio of 1:5.
All DC lines are filtered by a series of low-pass filters (LPFs) and thermalized to the refrigerator stages to minimize noise temperature.
Among the 17 DC lines, 11 are connected directly to the device, while the remaining lines are combined with microwave transmission lines and the readout tank circuit for qubit control and readout, respectively.
Baseband pulses are generated at room temperature using arbitrary waveform generators (AWGs) and transmitted through the qubit drive lines, each with a total attenuation of $-33$~dB.
The drive lines are combined with DC lines by on-board bias-tees and connected to gates $\text{P}_k$ and $\text{B}_l$ ($k=0,1,2$; $l=1,2$).
The readout line consists of an radio-frequency (rf) reflectometry circuit~\cite{connors2020rapid} incorporating a tank circuit and a DC bias line.
A local oscillator (LO) signal, controlled by an RF switch, is split by a three-port directional coupler to serve as both the probe and LO for signal down-conversion.
The probe tone is attenuated by approximately $-73$~dB and passes through a directional coupler at the mixing chamber.
One branch of the probe signal is delivered to the tank circuit—an on-board impedance-matching RLC network—connected to the ohmic contact $\text{S}_2$ of the single-electron transistor (SET).
Changes in the SET channel conductance result in a large on–off ratio in the reflection coefficient at the resonance frequency of the combined circuit.
The reflected signal, carrying information about the load reflection coefficient at the coupling port of the directional coupler, is first amplified by a low-noise amplifier mounted at the 1.5~K stage.
After passing through a high-pass filter (HPF) and an additional room-temperature amplifier, the signal is down-converted by mixing with a phase-shifted LO using a mixer.
The intermediate-frequency (IF) signal is subsequently filtered by two LPFs and digitized by an analog-to-digital converter (ADC).
A complete list of electronic components used in the setup is provided in Table~\ref{tab:S1_instrument}.

\section{Characterizing distortions of tunneling gates}

\begin{figure}[ht]
\centering
{\includegraphics[width=0.45 \textwidth]{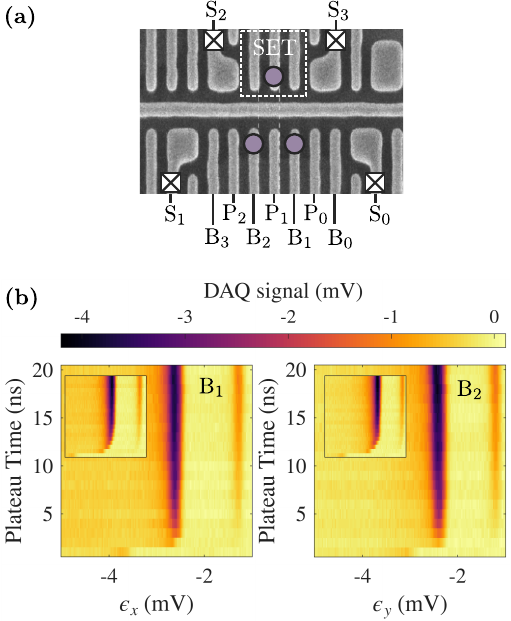}}
\caption{
(a) SEM image of a triple quantum-dot device that is similar to the one used in this experiment. 
A double quantum dot is tuned up under $\text{B}_1$ and $\text{B}_2$, where $\text{P}_1$ controls the inter-dot tunnel coupling. 
The charge state is sensed and read out by the single-electron transistor on the sensor side. 
(b) The DAPS results with and without pre-distortion corrections are obtained by individually pulsing gate $\text{B}_1$ and $\text{B}_2$. 
}
\label{figS:S2_calibrate_B1_B2}
\end{figure}

A double quantum-dot is tuned up under gates $\text{B}_1$ and $\text{B}_2$ for characterizing and correcting distortions, shown in Fig.~\ref{figS:S2_calibrate_B1_B2}(a).
The DQD is tuned up at the $(0,1)$-$(1,0)$ transition with a large inter-dot tunnel coupling.
The DAPS results with pre-distortions on gates $\text{B}_1$ and $\text{B}_2$ are illustrated in Fig.~\ref{figS:S2_calibrate_B1_B2}(b). 
The insets are the corresponding DAPS results without pre-distortion, which are used for calibrating the digital filters. 

\section{Digital filter design}
\subsection{Finite impulse response filters}

In realistic experimental setups, electrical distortions arise from multiple sources such as cable attenuation, impedance mismatch, filtering, and coaxial lines.
These effects can be accurately modeled as linear time-invariant (LTI) systems, since the response of each element depends linearly on the input signal and does not change with time.
The overall response of the entire signal chain can therefore be represented by a single, effective LTI system whose behavior is fully characterized by a discretized impulse response function $h[n]$ under a deterministic sampling rate.
In this framework, correcting pulse distortions is mathematically equivalent to inverting the LTI system via an appropriately designed digital filter.
A LTI system maps a discrete input signal $x[n]$ to an output $y[n]$ through convolution with its impulse response:
\begin{align}
    y[n] &= h[n]*x[n]\\ 
        &= \sum_{k=0}^N b_k x[n-k],
\end{align}
where $h[k]$ is the impulse response of length $N$ and the operator “$*$” denotes discrete convolution.
In digital implementation, such systems are realized as finite impulse response (FIR) filters, which are non-recursive and depend only on the present and a finite number of past samples of the input signal.
The impulse response ${h[k]}$ is equivalently represented by a set of discrete filter coefficients ${b_k}$:
\begin{equation}
    h[n] = \left\{ \begin{array}{cc}
        b_n, &  0\leq n \leq N \\
        0, & \text{otherwise}
    \end{array} \right. .
\end{equation}
To correct distortions, one seeks an inverse filter whose impulse response $h_{\mathrm{inv}}[n]$ satisfies
\begin{equation}
    (h_{\mathrm{inv}} * h)[n] \approx \delta[n],
\end{equation}
where $\delta[n]$ is the discrete Dirac delta function.
This ensures that the overall transfer function in $z$-domain $H_{\mathrm{inv}}(z)H(z)\approx 1$ across the signal bandwidth, effectively flattening the amplitude and phase response of the system.

The impulse response of the experimental setup can be obtained by measuring its step response, i.e., the system’s output when a digital step signal is applied.
Differentiating this transient response over a finite time window yields $h[n]$ matching the instrument sampling rate.
The corresponding convolution matrix can be constructed as 
\begin{equation}
    \mathcal{H} = \left[
    \begin{array}{cccccc}
         h[0]& 0 & 0 & \dots & 0 & 0 \\
         h[1]& h[0] & 0 & \cdots & 0 & 0\\
         h[2]& h[1] & h[0] & \cdots & 0 & 0\\
         \vdots & \vdots & \vdots & & \vdots & \vdots\\
         h[N] & h[N-1] & h[N-2] & \cdots & h[1] & h[0] \\
         0 & h[N] & h[N-1] & \cdots & h[2] & h[1] \\
         \vdots & \vdots & \vdots & & \vdots & \vdots\\
         0 & 0 & 0 & \cdots & h[N] & h[N-1] \\
         0 & 0 & 0 & \cdots & 0 & h[N]
    \end{array}
    \right].
\end{equation}
The inverse of this matrix, $\mathcal{H}^{-1}$, represents the operator that pre-distorts the input waveform to compensate for the measured system response.
In practice, directly inverting $\mathcal{H}$ can lead to overshooting artifacts at short timescales.
To mitigate this, we apply Tikhonov regularization~\cite{tikhonov1963solution, hansen2010discrete, vogel2002computational} to the inverse:
\begin{align}\label{eqS:inverse_filter_regu}
   \mathcal{H}_{\mathrm{inv}} = \mathcal{H}^{-1} + \lambda^2 I,
\end{align}
where $\lambda$ is the regularization parameter and $I$ is the identity matrix.
The regularization suppresses large corrections in the inverse filter that arise from poorly conditioned components of $\mathcal{H}$.
For all filters calibrated in this work, we use $\lambda=1$.
As demonstrated in Sec.~\ref{SecS:filter_optimization}, $\lambda$ can also be optimized automatically via a gradient descent algorithm, providing a fully automated approach to overshoot mitigation.
In the DAPS cryoscope experiment, square pulses are generated at a sampling rate $f_s$ and sent to the double quantum-dot (DQD).
By measuring the time-resolved peak positions of the spectroscopy signal, one can extract the effective voltage amplitude that brings the system to the avoided crossing at each plateau time.
This measurement directly yields information about the step response, from which a pre-distortion FIR filter can be constructed to compensate the effective waveform at the device level.

\begin{figure}[ht]
\centering
{\includegraphics[width=0.7 \textwidth]{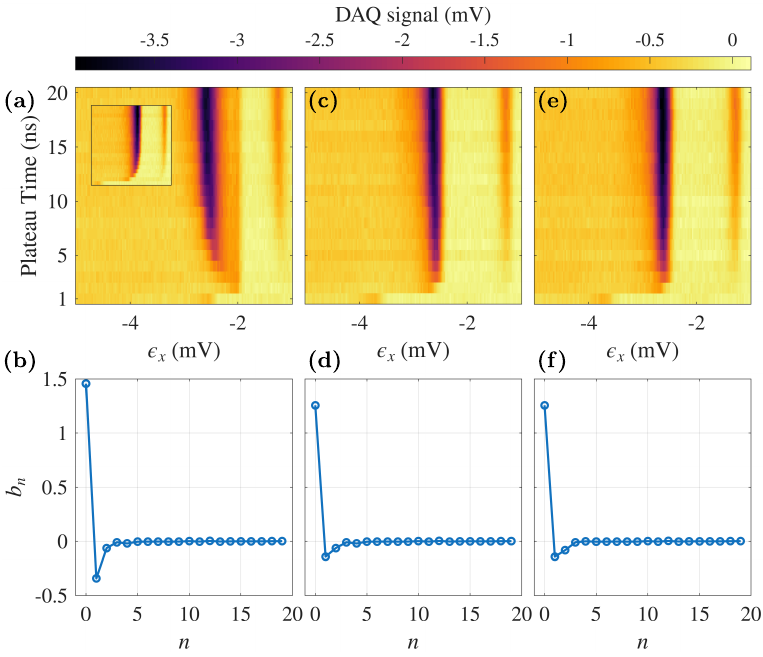}}
\caption{
(a) B$_1$-B$_2$ DQD one-electron DAPS along the B$_1$ axis with pre-distortion via FIR filter before overshoot adjustments. 
The initial FIR filter, shown in (b), is calibrated from the DAPS result in the inset.
The overshoot introduces copies of the DAPS lines and causes them to move back and forth.
(c) DAPS with pre-distortion after the first manual adjustment to $b_0$ and $b_1$.
The corresponding filter coefficients are shown in (d).
(e) DAPS with pre-distortions after the second manual adjustment to $b_2$ and $b_4$.
The corresponding filter coefficients are shown in (f).
}
\label{figS:S3_overshooting}
\end{figure}

\subsection{Overshooting}

In the experiment, however, we observe that all calibrated filters introduce overshooting at short timescales, as mentioned in the main text and explained further in the next section.
Figure~\ref{figS:S3_overshooting}(a) shows an example of the measured overshoot of a pre-distorted square pulse applied to gate B$_1$, with the corresponding uncorrected DAPS result shown in the inset and calibrated FIR filter $\{b_n\}$ in Fig.~\ref{figS:S3_overshooting}(b).
Both spectroscopy lines, corresponding to the transitions $|g_-\rangle\leftrightarrow|e_-\rangle$ (right line) and $|g_-\rangle\leftrightarrow|e_+\rangle$ (left line), bend in opposite directions relative to the result obtained without pre-distortion.
For avoided crossings with stronger coupling strength, additional copies of these spectroscopy lines appear, a clear signature of overshooting.
Physically, this overshoot arises because the DAPS-extracted impulse response overestimates the effective time constant of the drive line at short timescales, as discussed in Sec.~\ref{SecS:model_dynamics}.
The Tikhonov regularization term $\lambda^2 I$ in Eq.~(\ref{eqS:inverse_filter_regu}) primarily modifies the leading filter coefficient $b_0$, effectively shortening this time constant; for an appropriate choice of $\lambda$, the overestimation is corrected and the overshoot is suppressed.
Here, we apply manual adjustments to the FIR filter coefficients to mitigate the overshoot.
As demonstrated in Sec.~\ref{SecS:filter_optimization}, such mitigation can also be implemented automatically by optimizing $\lambda$ via a gradient descent algorithm.
Figure~\ref{figS:S3_overshooting}(c) shows the result after the adjustment
\begin{equation}
    b'_n = 
    \begin{cases}
        b_n-0.2 & n=0 \\
        b_n+0.2 & n=1 \\
        b_n     & \text{otherwise}
    \end{cases},
\end{equation}
where the first-adjusted filter $\{b'_n\}$ is shown in Fig.~\ref{figS:S3_overshooting}(d).
Figure~\ref{figS:S3_overshooting}(e) shows the result after the adjustment
\begin{equation}
    b''_n = 
    \begin{cases}
        b'_n-0.02 & n=2 \\
        b'_n+0.02 & n=4 \\
        b'_n     & \text{otherwise}
    \end{cases},
\end{equation}
where the second-adjusted filter $\{b''_n\}$ is shown in Fig.~\ref{figS:S3_overshooting}(f).
Both results suggest that the overshoot is mitigated, at the expense of re-introducing undershoot in the first few nanoseconds. 
The adjusted filter in Fig.~\ref{figS:S3_overshooting}(d) are used in the following 4e-DAPS and ST$_0$ qubit experiments. 

\section{Signal-to-noise ratio}
\subsection{Model and dynamics}\label{SecS:model_dynamics}

\begin{figure}[ht]
\centering
{\includegraphics[width=1 \textwidth]{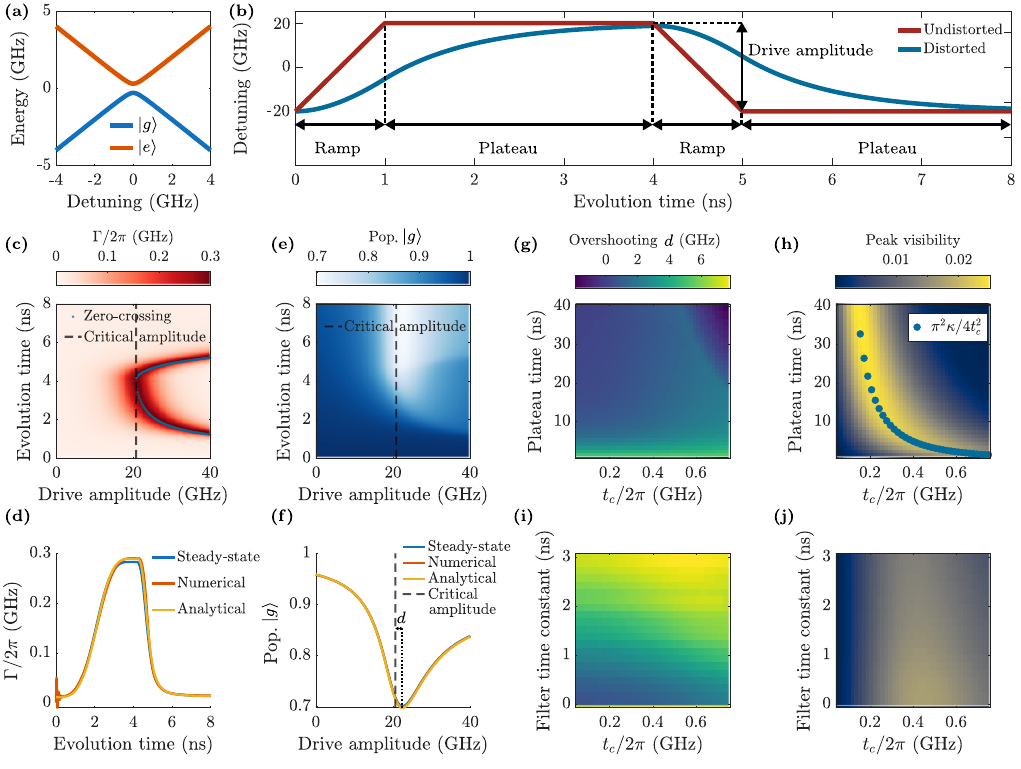}}
\caption{
(a) Energy-levels of the two-level tunneling model.
(b) Detuning pulse $\epsilon(t)$ with and without distortions.
(c) Analytical relaxation rate $\Gamma$ from Equation~(\ref{eqS:analytical_gamma_SM}) with a distorted detuning drive at different drive amplitudes and evolution times.
The simulated detuning pulse has a 3-ns plateau time and a 1-ns ramp time.
The critical amplitude is where highest point of the detuning pulse reaches the avoided crossing at $\epsilon=0$.
The system passes the avoided crossing twice when the drive amplitude is above the critical amplitude.
(d) Vertical trace at the critical amplitude in (c), plotted with both the steady-state and numerical solutions from Equation~(\ref{eqS:master_eq}).
(e) The corresponding $|g\rangle$-state population obtained by integrating the analytical relaxation rate in (c), assuming the system is initialized in $|g\rangle$.
(f) Horizontal traces from (e) at 8~ns. 
The overshoot factor $d$ is defined as the difference between the critical amplitude and the spectroscopy peak.
Simulated (g) overshoot factor $d$ and (h) peak visibility at different plateau times and $t_c$ with a fixed distortion filter time constant $\tau = 1$~ns.
Simulated (i) $d$ and (j) peak visibility with different distortion filter time constants and $t_c$ with a plateau time $5$~ns. 
}
\label{figS:S4_simulated_dynamics}
\end{figure}

To get a better understanding of the performance and working range of this cryoscope, we study Landau-Zener-type dynamics in a reduced two-level system.
For simplicity, we assume $\hbar=1$ in the following derivations and simulation results. 
Consider a two-level tunneling model with a static Hamiltonian
\begin{equation} \label{eqS:tls_H}
    H_0 = \frac{1}{2} \left( \epsilon\sigma_z + t_c \sigma_x \right),
\end{equation}
where $\epsilon$ is detuning and $t_c$ is transversal tunnel coupling. 
The energy levels around zero-detuning are illustrated in Fig.~\ref{figS:S4_simulated_dynamics}(a), where the ground and excited eigenstates are labeled by $|g\rangle$ and $|e\rangle$, with corresponding eigenvalues $\mp\sqrt{\epsilon^2+t_c^2}/2$, respectively.
Suppose that a detuning pulse $\epsilon(t)$ is applied to the system, which includes an excursion and return to an idling point. Example ideal and distorted pulse are shown in Fig.~\ref{figS:S4_simulated_dynamics}(b).  
We simulate the system dynamics with a Lindblad master equation with a Markovian detuning noise:
\begin{equation} \label{eqS:master_eq}
    \dot{\rho} = -i \left[ H_0 , \rho \right] + \frac{\kappa}{2} \mathcal{D}[\sigma_z] (\rho),
\end{equation}
where $\rho$ is density matrix of the system, $\kappa$ is the dephasing rate introduced by the detuning noise, and $\mathcal{D}$ is the dissipator. 
By defining the Bloch vector $u = 2\text{Re}(\rho_{01})$, $v = 2\text{Im}(\rho_{01})$, and $w=\rho_{11}-\rho_{00}$, Equation~(\ref{eqS:master_eq}) becomes a set of optical Bloch equations:
\begin{align} 
    \dot{y} &= M y - b^\intercal w, \label{eqS:OB_coherence} \\ 
    \dot{w} &= by. \label{eqS:OB_population} 
\end{align}
Here, $y = (u,v)^\intercal$, $b = (0,t_c)$, and the matrix $M$ and its inverse are given by
\begin{align}
    M &= \left[ 
    \begin{array}{cc}
         -\kappa & \epsilon \\
         -\epsilon & -\kappa  
    \end{array}
    \right],\\
    M^{-1} &= \frac{1}{D} \left[ 
    \begin{array}{cc}
         -\kappa & -\epsilon \\
         \epsilon & -\kappa  
    \end{array}
    \right],
\end{align}
where $D = \epsilon^2 + \kappa^2$.
In the limit of large $\kappa$, the solutions of Equations~(\ref{eqS:OB_coherence})-(\ref{eqS:OB_population}) can be approximated via asymptotic expansion. 
For $\kappa \gg t_c$, and at long times, i.e., $t\gg \kappa^{-1}$, the lost of coherence is much faster than the relaxation on $w$. 
Therefore, we take the zeroth-order solution $y_0$ to be such that $\dot{y_0} = 0$ in Equation~(\ref{eqS:OB_coherence}), and we obtain
\begin{equation}
    y_0 = M^{-1} b^\intercal w = -\frac{t_c w}{D} \left(
    \begin{array}{c}
         \epsilon\\
         \kappa
    \end{array}
    \right).
\end{equation}
Inserting this into Equation~(\ref{eqS:OB_population}) leads to the well-known steady-state solution of the relaxation rate
\begin{equation}
    \Gamma_{\text{st}} =  \frac{\kappa t_c^2}{D}.
\end{equation}
The first-order solution is evaluated within the time scale $t\sim \kappa^{-1}$ before the system completely loses its coherence. 
We assume the first-order solution $y_1$ and substitute the total solution $y = y_0 + y_1$ into Equation~(\ref{eqS:OB_coherence}).
Using $My_0 - b^\intercal w = 0$, the equation reduces to
\begin{equation} \label{eqS:first_order_y}
    \dot{y_0} + \dot{y_1} = M y_1.
\end{equation}
When $t\sim \kappa^{-1}$, the first-order correction $y_1$ can be treated as quasi-static ($\dot{y_1} = 0$), while $\dot{y_0}\neq 0$.
In this approximation, Equation~(\ref{eqS:first_order_y}) yields the solution of $y_1$ as
\begin{equation}
    y_1 = M^{-1} \dot{y_0} = \frac{t_c}{D^2} \left(
    \begin{array}{c}
         \kappa \dot{\epsilon} w\\
          D\dot{w} - \epsilon\dot{\epsilon}w
    \end{array}
    \right).
\end{equation}
Therefore, the asymptotic solution of $y$ up to the first-order leads to a effective relaxation rate
\begin{equation} \label{eqS:analytical_gamma_SM}
    \Gamma = - \frac{\dot{w}}{w} = \frac{t_c^2 (\kappa D + \epsilon\dot{\epsilon})}{D(D - t_c^2)}.
\end{equation}

To validate the asymptotic solution, we numerically simulate the system initialized in the ground state $|g\rangle$ at a detuning of $\epsilon/2\pi = -20~\mathrm{GHz}$, using parameters $t_c/2\pi = 0.3~\mathrm{GHz}$ and $\kappa/2\pi = 2~\mathrm{GHz}$.
Figures~\ref{figS:S4_simulated_dynamics}(c) and (e) show the analytical results for the relaxation rate and the population in $|g\rangle$ under a distorted detuning drive, plotted against varying evolution time and detuning drive amplitude.
A fixed ramp time of 1~ns and plateau time of 3~ns are used in these two simulations. The range of the $y$ axis corresponds to the full 8~ns length of the simulated detuning pulse.
Distortions in the detuning waveform are modeled by applying an infinite-impulse response (IIR) low-pass filter with a time constant $\tau=1$~ns.
For a given waveform, there is a critical drive amplitude such that the maximum value of the pulse  reaches the avoided crossing at $\epsilon = 0$.
Near this point, the relaxation rate approaches its maximum because longitudinal dephasing is converted into effective transverse relaxation in the eigenbasis as the two energy levels become resonant. This critical amplitude for each plateau time corresponds to the actual amplitude required for the detuning to reach the avoided crossing at the end of the plateau, including the effects from distortions.
Figure~\ref{figS:S4_simulated_dynamics}(d) plots a line cut at this critical drive amplitude, where the analytical relaxation rate in Equation~(\ref{eqS:analytical_gamma_SM}) agrees well with the numerical evaluation obtained from simulating Equation~(\ref{eqS:master_eq}).
The steady-state solution, however, shows a slight under-estimation of the relaxation rate. 

The analysis discussed in the main text assumes that the critical amplitude is also the amplitude that gives the largest population change. 
However, as shown in Figs.~\ref{figS:S4_simulated_dynamics}(e) and (d), the drive amplitude giving the largest population difference after the entire pulse differs from the critical amplitude by an amount that we denote as $d$.
This mismatch manifests experimentally as an overshoot induced by the pre-distortion filter when the impulse response extracted from DAPS is used without adjustment.

The origin of this mismatch is intrinsic: the measured population represents a time-integrated response of the instantaneous relaxation rate. 
For a waveform with finite ramp edges and non-monotonic rates, where the detuning increases and then decreases, the cumulative relaxation is maximized at amplitudes slightly above the critical value, even in the absence of distortions.
This overshooting effect is suppressed at long plateau times so that the overall dynamics are dominated by steady-state relaxation rather than the contributions from the transient of the time-dependent detuning.

To further understand the performance of the DAPS cryoscope, we extract the overshoot factor $d$ from numerical simulations at different tunnel couplings $t_c$, distortion filter time constants $\tau$, and plateau times.
Here, we consider fixed dephasing rate $\kappa/2\pi = 2$~GHz and ramp time 1~ns in all detuning pulses.
Figures~\ref{figS:S4_simulated_dynamics}(g) and (i) show $d$ at different parameters with fixed $\tau= 1$~ns and plateau time 5~ns, respectively.
The overshoot factor weakly depends on $t_c$ and is significantly suppressed at longer plateau time and shorter filter time. 
This result matches the experimental observation that overshooting happens at short times and is dominated by the first few terms in the impulse response function $h[n]$.

\subsection{Effect of $t_c$ on peak visibility}

We define the DAPS peak visibility as
\begin{equation}
    \mathcal{V} = \frac{w_{\text{pk}}}{\text{Peak width}},
\end{equation}
where $w_\text{pk}$ is the magnitude of DAPS peaks signal.
Figure~\ref{figS:S4_simulated_dynamics}(h) and (j) shows the simulated peak visibility at different parameters with fixed $\tau = 1$~ns and plateau time 5~ns, respectively. 
At short plateau times, the peak visibility is high at large $t_c$.
This result matches the experimental observation that the spectroscopy peaks appear earlier in time when we increase the tunnel coupling.
The high visibility region moves towards small $t_c$ as the peaks are broadened at large $t_c$ and large plateau time.
The plateau time with high visibility can be estimated by a empirical formula $\pi^2\kappa /4t_c^3$, as shown in Fig.~\ref{figS:S4_simulated_dynamics}(h).
For a fixed plateau time, the peak visibility shows weak dependency on the distortion time constant $\tau$.

\subsection{Predicting overshoot factor at 1~ns for P$_0$}\label{SecS:predicting_overshoot}

\begin{figure}[ht]
\centering
{\includegraphics[width=\textwidth]{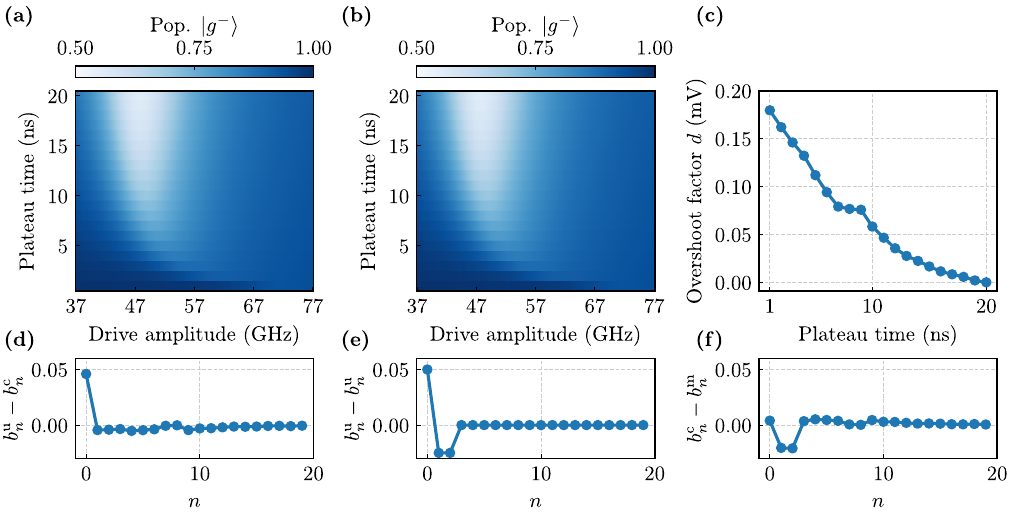}}
\caption{
Predicting the overshoot correction from the measured pulse waveform. 
(a) Numerical and (b) analytical simulations of DAPS using the detuning pulse waveform measured in Fig.~2. 
The agreement between the two validates the analytical framework for realistic pulse shapes. 
(c) Overshoot factor $d$ extracted from the analytical result as a function of plateau time. 
The overshoot is largest at short plateau times and decays as the dynamics become dominated by steady-state relaxation. 
(d) Difference between uncorrected and overshoot-corrected filter coefficients $b_n^\text{u}-b_n^\text{c}$. 
(e) Difference between uncorrected and independently manually adjusted filter coefficients $b_n^\text{u}-b_n^\text{m}$. 
(f) Difference between overshoot-corrected and manually adjusted filter coefficients $b_n^\text{c}-b_n^\text{m}$. 
}
\label{figS:S5_real_wf}
\end{figure}

In this section, we use the framework developed in Sec.~\ref{SecS:model_dynamics} to estimate the overshoot factor for the waveform measured from DAPS data in Fig.~2. 
For simplicity, we ignore the extra states $|g_+\rangle$ and $|e_-\rangle$, retaining only $|g_-\rangle$ and $|e_+\rangle$, which form the relevant avoided crossing for measuring the response.
Using the same simulation code as in Fig.~\ref{figS:S4_simulated_dynamics}, we numerically simulate the system initialized in the grounded state $|g_-\rangle$ at a detuning $\epsilon_x/2\pi=-47.23$~GHz, with parameter $t_c/2\pi=0.2$~GHz, $\kappa/2\pi = 2$~GHz, and lever arm $\alpha=0.1$~eV/V. 
We note that the overshoot factor $d$ depends weakly on the tunnel coupling $t_c$~[Fig.~\ref{figS:S4_simulated_dynamics}(g)].
An approximate value of $t_c$, which can be constrained from the DAPS peak width and signal level, is therefore sufficient for estimating the overshoot correction.
Figures~\ref{figS:S5_real_wf}(a) and (b) show the simulated DAPS results using numerical and analytical approaches, respectively, and they agree with each other, validating the analytical framework for realistic pulse shapes. 
Using the analytical result, we extract the overshoot factor $d$ at different plateau times [Fig.~\ref{figS:S5_real_wf}(c)].
As expected from the analysis in Sec.~\ref{SecS:model_dynamics}, $d$ is largest at short plateau times, precisely where pulse distortions are most severe, and decays at longer plateau times where the dynamics are dominated by steady-state relaxation.
To verify the predicted overshoot factor $d$ matches the experimental observation, we correct the pre-distortion FIR filter by compensating the overshoot voltage on the measured detuning response. 
We denote the uncorrected filter (which introduces the overshoot) as $\{b_n^\text{u}\}$, the overshoot-corrected filter as $\{b_n^\text{c}\}$, and the independently manually adjusted filter as $\{b_n^\text{m}\}$ for the P$_0$ gate.
The manually adjusted filter $\{b_n^\text{m}\}$ is the one employed in all experimental results in the main text and was obtained without reference to the overshoot prediction.
Figures~\ref{figS:S5_real_wf}(d), (e) and (f) show the differences between $b_n^\text{u}-b_n^\text{c}$, $b_n^\text{u}-b_n^\text{m}$, and $b_n^\text{c}-b_n^\text{m}$, respectively.
Comparing panels (d) and (e), the corrections predicted by the overshoot compensation closely reproduce the structure of the manual adjustment, which both concentrated at $n=0$ to reduce the overshooting voltage at 1~ns.
Most importantly, panel (f) shows that the overshoot-corrected and manually adjusted filters are in closest agreement at $n=0$, where the overshoot is most dominant.
The residual differences at $n=1$ and $n=2$ arise because the manual correction redistributes the total correction weight across only the first three coefficients~($n=0,1,2$), where the overshoot-corrected filter distributes it across all twenty coefficients. 
Despite this difference in correction strategy, both filters achieve comparable overshoot mitigation at the leading timescale.
This result demonstrates that the overshoot is quantitatively predictable from the DAPS-measured step response using the analytical framework, confirming that the manual adjustments employed in this work are physically grounded.
While the analytical prediction requires approximate knowledge of device parameters such as $t_c$, the automated filter optimization demonstrated in Sec.~\ref{SecS:filter_optimization} eliminates this requirement entirely, providing a parameter-free path to overshoot mitigation.

\subsection{Estimating $\kappa$ from $1/f$ detuning noise}

\begin{figure}[ht]
\centering
{\includegraphics[width=0.5 \textwidth]{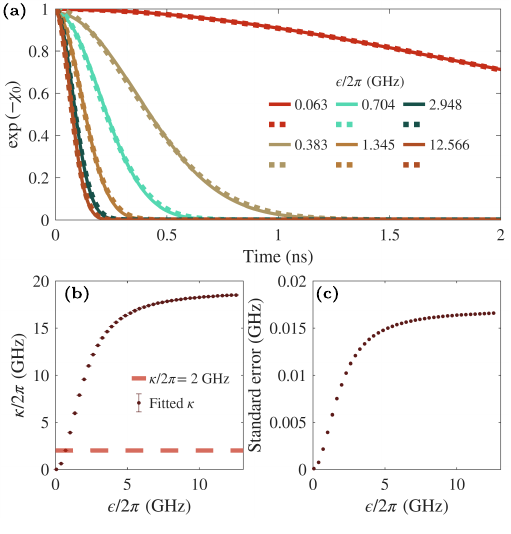}}
\caption{
(a) Simulation of the decay transients associated with the coherence function at different detunings. 
(b) Fitted $\kappa$ at different detunings.
(c) Corresponding standard errors of the fitted $\kappa$.
}
\label{figS:S6_estimate_kappa}
\end{figure}

The detuning-noise-induced dephasing rate $\kappa/2\pi = 2$~GHz, employed in all simulations, applies to the case of a Markovian environment, strictly speaking.
Realistically, however, electrical noise is highly non-Markovian, with a characteristic $1/f^\beta$ noise power spectral density (PSD).
By fitting the decay associated with the coherence function of a two-level system, we extract an estimation of the dephasing rate $\kappa$. 
Consider the Hamiltonian in Equation~(\ref{eqS:tls_H}) under a detuning noise $\delta \epsilon$:
\begin{equation}
    H_0 = \frac{1}{2} \left[ (\epsilon + \delta \epsilon ) \sigma_z + t_c \sigma_x \right],
\end{equation}
By defining the qubit transition frequency as $\omega_q = \sqrt{\epsilon^2 + t_c^2}$, its coherence function associated with a free-induction decay is~\cite{bylander2011noise} 
\begin{equation}
    \chi_0(t) = t^2 \left( \frac{\partial \omega_q}{\partial \epsilon} \right)^2 \int_0^\infty g_0 (\omega, t) \, S_\epsilon (\omega) \, d\omega.
\end{equation}
Here, $S_\epsilon(\omega)$ is the PSD of the detuning noise, $t$ is the evolution time, and
\begin{equation}
    g_0(\omega, t) = \left(\frac{\sin (\omega t /2 )}{\omega t /2} \right)^2.
\end{equation}
We assume the PSD has the form
\begin{equation}
    S_\epsilon (\omega) = \frac{A^2_{1\text{Hz}}}{f},
\end{equation}
where the measured PSD amplitude at 1~Hz in Si/SiGe quantum dots is $A_{1\text{Hz}} \approx 1\mu\text{eV}$~\cite{connors2019low}.
With this assumption, the numerically-evaluated decay associated with the qubit coherence function $\exp[-\chi_0 (t)]$ is shown in Fig.~\ref{figS:S6_estimate_kappa}(a) at different detunings $\epsilon$, with $t_c/2\pi = 0.2$~GHz.
The coherence is preserved near the zero-detuning because the longitudinal detuning noise becomes transversal in the energy basis, causing relaxation instead of dephasing.
By fitting the decay curves to a function of the form $\exp(-\kappa t^2)$, we extract the dephasing rate  [Figs.~\ref{figS:S6_estimate_kappa}(b) and (c)]. 
Based on the extracted dephasing rates, we choose  $\kappa/2\pi = 2$~GHz for our simulations.

\subsection{Simplified case}

In this section we derive Equations~(2) and (3) in the main text. 
We assume $\kappa \gg t_c$ and work in the steady-state (incoherent) limit.
Starting from the two-level Hamiltonian in Equation~(\ref{eqS:tls_H}) and assuming a negligible orbital relaxation $\Gamma_\downarrow~(\Gamma_\uparrow)$, the dynamics of the upper-branch population are governed by a rate equation. 
To match the step response for extracting filter coefficients, we redefine $w$ such that it increases from 0 to 1 during the evolution.
Redefining $w \equiv p_e - p_g + 1 = 2p_e$ as the probability for an energy measurement of the system to yield the value corresponding to the $\ket{e}$ eigenstate, we obtain
\begin{equation}\label{eqS:simplified_case_evo_eq}
    \dot{w}(t) = -\Gamma_\mathrm{st}(t)\, \left[w(t)-1\right],
\end{equation}
where the steady-state relaxation rate is
\begin{equation}
    \Gamma_\mathrm{st}(t) =
    \frac{\kappa t_c^2}{\kappa^2 + [\epsilon(t)-\epsilon_0]^2}.
\end{equation}
Here $\epsilon_0$ denotes the detuning at which the two levels are on resonance.
We consider a distorted step detuning pulse of the form $\epsilon(t)=\epsilon s(t)$, where $s(t)$ is the step response of the control line. We assume $\epsilon(t)$ increases monotonically in time, $\dot{\epsilon}(t)>0$, and that the system is initialized in the ground state, i.e., $p_g=1$ and $p_e=0$, so that $w(0)=0$.
Equation~(\ref{eqS:simplified_case_evo_eq}) then has the formal solution
\begin{equation}
    w(t) = 1-\exp\!\left[-\int_0^t \Gamma(x)\, dx\right].
\end{equation}
Since $\epsilon(t)$ is monotonic, we may change integration variables from time to detuning by defining $\varepsilon=\epsilon(x)$. Using
$d\varepsilon=\dot{\epsilon}(x)\, dx$ and $x=s^{-1}(\varepsilon/\epsilon)$, we obtain
\begin{equation}
    w(t,\epsilon)= 1-
    \exp\!\left[
    -\int_0^{\epsilon s(t)}
    \frac{\Gamma(\varepsilon)}
    {\dot{\epsilon}\!\left(s^{-1}(\varepsilon/\epsilon)\right)}
    \, d\varepsilon
    \right],
\end{equation}
which is Equation~(2) in the main text.
When the resonance is narrow compared to the scale over which $\epsilon(t)$ varies, we may approximate
$\Gamma(\varepsilon)\approx \pi t_c^2\delta(\varepsilon-\epsilon_0)$.
Evaluating the integral then yields
\begin{equation}
    w(t,\epsilon)\approx
    \begin{cases}
        1-\exp\!\left[-\dfrac{\pi t_c^2}{\dot{\epsilon}(t')}\right], & t'\le t,\\[6pt]
        0, & \textnormal{otherwise},
    \end{cases}
\end{equation}
where $t'=s^{-1}(\epsilon_0/\epsilon)$ denotes the time at which the distorted detuning pulse crosses the resonance.
For a fixed pulse duration $t$, the condition $t' \le t$ indicates that the system traverses the avoided crossing during the evolution. 
In this case, incoherent mixing between the two diabatic states occurs, resulting in an exponential approach toward the mixed state. In contrast, when $t' > t$, the system does not reach the avoided crossing within the pulse duration. 
As we neglect orbital relaxation in this calculation, the populations remain unchanged and $w(t,\epsilon)\approx 0$.
This result corresponds to Eq.~(3) in the main text.

\subsection{Two-level fluctuators}

\begin{figure}[ht]
\centering
{\includegraphics[width=0.45 \textwidth, trim={0 0cm 0cm 0}, clip]{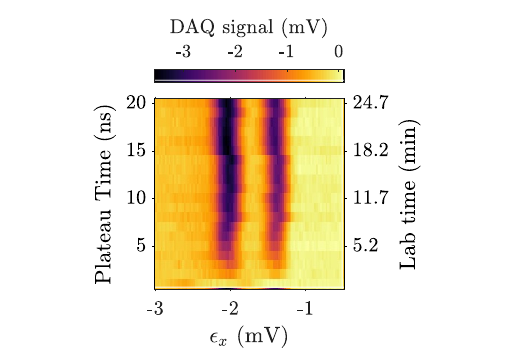}}
\caption{
DAPS data for the DQD under gate $\text{P}_0$ and $\text{P}_1$ with pre-distortions. The DQD is likely coupled to a few slow-switching TLFs, resulting in shifting spectroscopy peaks. 
}
\label{figS:S7_TLF}
\end{figure}

The performance of our cryoscope is also limited by individual low-frequency charged two-level fluctuators (TLFs) that couple to the DQD.
Recent research shows that TLFs in Si/SiGe spin qubits can have a characteristic switching time range between 10~ms and 20~s~\cite{ye2024characterization, yoneda2023noise}.
To resolve such random telegraph noise in the DAPS measurements without averaging it out, we perform consecutive DAPS measurements row-by-row, shown in Fig.~\ref{figS:S7_TLF}, where the lab time is indicated by the axis on the right. 
Shifting spectroscopy peaks are observed due to TLFs switching between different charge states, which significantly reduce the accuracy for extracting peak positions.

\section{$|S\rangle \leftrightarrow |T_-\rangle$ transition}

\begin{figure}[ht]
\centering
{\includegraphics[width=0.45 \textwidth]{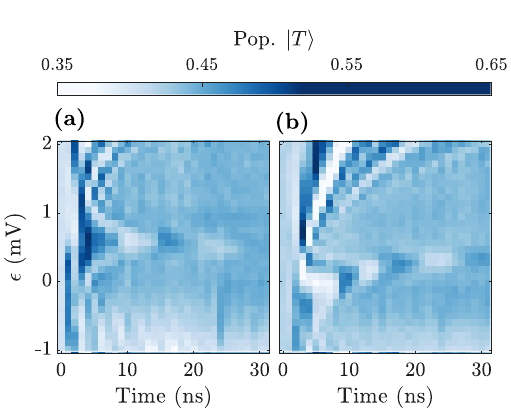}}
\caption{
$|S\rangle \leftrightarrow |T_-\rangle$ transition (a) with pre-distortions, and (b) without pre-distortions on all finger gates.
}
\label{figS:S8_STm}
\end{figure}

To further visualize the effect of our pre-distortion filter, we probe the $|S\rangle \leftrightarrow |T_-\rangle$ transition by pulsing both plunger gates along the detuning axis.
A singlet is prepared by adiabatic ramping from charge state $(0,4)$ to $(1,3)$.
Then, the system is quickly pulsed to a detuning point and simultaneously a tunneling gate pulse is applied with amplitude 75~mV.
This large tunneling pulse modifies the confinement potential of both dots such that the coupling strength between $|S\rangle$ and $|T_-\rangle$, primarily resulting from spin-orbit coupling, is enhanced.
Then, the system is adiabatically ramped back to the measurement point inside the Pauli spin blockade region to perform a readout.
Figure~\ref{figS:S8_STm}(a) and (b) show a ``straight'' Chevron pattern when pre-distortions are applied to all gates, while a ``curved'' Chevron pattern appears without pre-distortions. 

\section{Short-time Fourier analysis of frequency chirp}

On top of the sliding-window fit algorithm shown in the main text, we use an alternate method to determine the presence of a chirp in the signal. We use a short-time Fourier transform (STFT) in which a conventional Fourier transform is applied as a sliding moving window to the signal. This is formally computed as,
\begin{equation}
    \mathrm{SFTF}\{x(t)\}(\tau,\omega) = X(\tau,\omega) = \int_{-\infty}^{\infty} x(t) \, \omega(t-\tau)e^{-i\omega t} \, dt,
\end{equation}
where the objective signal is $x(t)$, the window function is $\omega(t-\tau)$.
This transform provides the frequency components at different times for the given signal. 
The presence of a chirp will be reflected as a slope in the spectrogram, with positive chirp corresponding to positive slope and vice versa.

In order to further characterize any slope, we apply a Radon transform. This can be understood as extracting the uni-dimensional projections of the time-frequency intensity distribution from the spectrogram along different angles with respect to the time axis. 
This way, an intensity distribution with a slope at certain angle corresponds to high Radon response at the same angle. 
The definition of the Radon transform is
\begin{equation}
    R\{f(x,y)\}(x',\theta) = \int_{-\infty}^{\infty} f(x' \cos\theta-y'\sin\theta, x' \sin\theta + y'\cos\theta) \,  dy',
\end{equation}
where $\theta$ is the angle of the projection and $x'$ is the variable of the projection. 
Notice that this definition implies that the variables of the function have the same units. 
For this case, we consider arbitrary units and focus solely on the Radon response intensity distribution. 
This way a change of the chirp will be directly reflected in how the Radon response is distributed. 
If the signal has no chirp, we expect a strong Radon response at $\theta=90^{\circ}$. If the signal has a chirp, the maximum of the signal should occur away from $90^{\circ}$.

\begin{figure}[ht]
    \centering
    \includegraphics[]{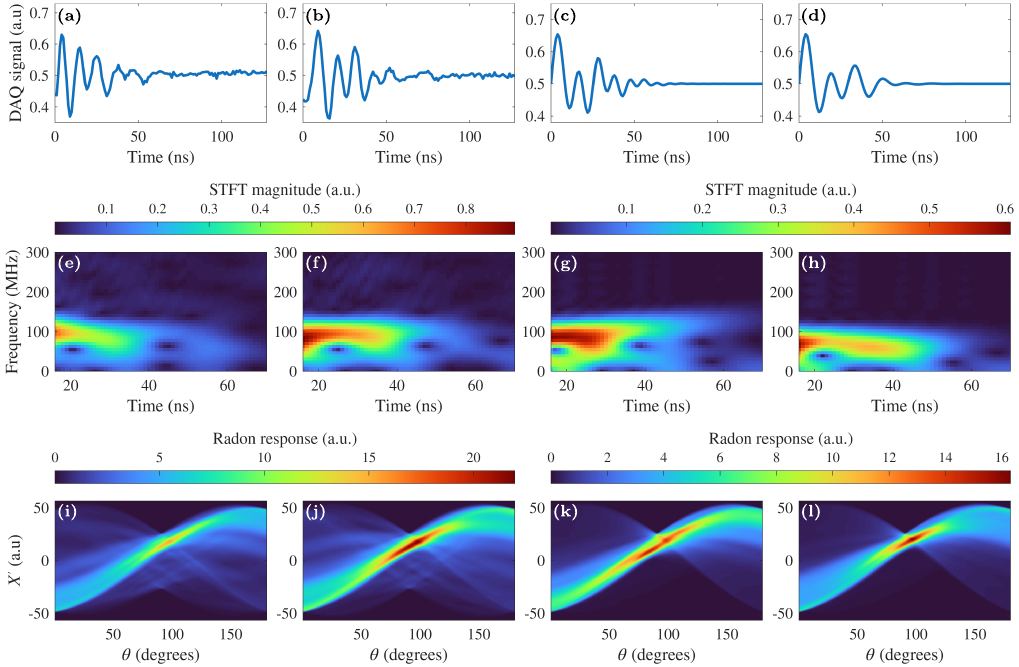}
    \caption{Signal analysis of singlet-triplet exchange oscillations. (a)-(d) Pulsed exchange oscillations at $\epsilon=2$mV. (e)-(d) Spectrograms of exchange oscillation time traces. (i)-(l) Radon transform of spectrograms. Columns: experimental data with pre-distortion, experimental data without pre-distortion, simulated data with chirp applied, simulated data without chirp applied.}
    \label{fig: S9_STFT}
\end{figure}

We apply this method to singlet-triplet exchange oscillations with and without pre-distortion applied as shown in Fig.~\ref{fig: S9_STFT}. 
Given that the spectrograms have a non-trivial time-frequency intensity distribution, we use simulated time trace signals for comparison. 
The model for the simulation is as follows,
\begin{equation}
   s(t) =  \left[A_{1} \sin(2\pi B_{1} t + 2\pi C^{2} t^{2}) + A_{2}\sin(2\pi B_{2}t)\right] \exp(-t^{2}/t_{0}^{2}) + D.
\end{equation}
The coefficients used for Fig.~\ref{fig: S9_STFT}(c)-(d) are shown in Table~\ref{tab: params S8}.

\begin{table}[]
\begin{ruledtabular}
\begin{tabular}{lcccccccc}
  & $A_{1}$ (a.u) & $A_{2}$ (a.u) & $B_{1}$ (MHz) & $B_{2}$ (MHz) & $C$ (MHz) & $D$ (a.u)  & $t_{0}$ (ns) \\\colrule
Params. for Fig.~\ref{fig: S9_STFT}(c) & $0.1044$  &  $0.0610$  & $65.1$  & $40.7$  &$22.599$  & 0.5  & $35.394$   \\
Params. for Fig.~\ref{fig: S9_STFT}(d) & $0.1044$  &  $0.0610$  & $65.1$  & $40.7$  & 0 &  0.5  & $35.394$  
\end{tabular}
\end{ruledtabular}
\caption{Parameters for simulated singlet-triplet exchange oscillation time traces.}
\label{tab: params S8}
\end{table}

By comparing the columns of Fig.~\ref{fig: S9_STFT}, notice first for the simulation results that the spectrogram captures how the chirp affects the slope of the time-frequency intensity distribution, which is reflected by a higher Radon response around $\theta = 60^{\circ}$. 
Looking at the experimental results, we see that the pre-distortion filters move the maximum Radon response closer to $\theta=90^{\circ}$, suggesting that the chirp is reduced. 

\section{Mitigating overshoot via filter optimization}\label{SecS:filter_optimization}

\begin{figure}[ht]
\centering
{\includegraphics[width=1 \textwidth]{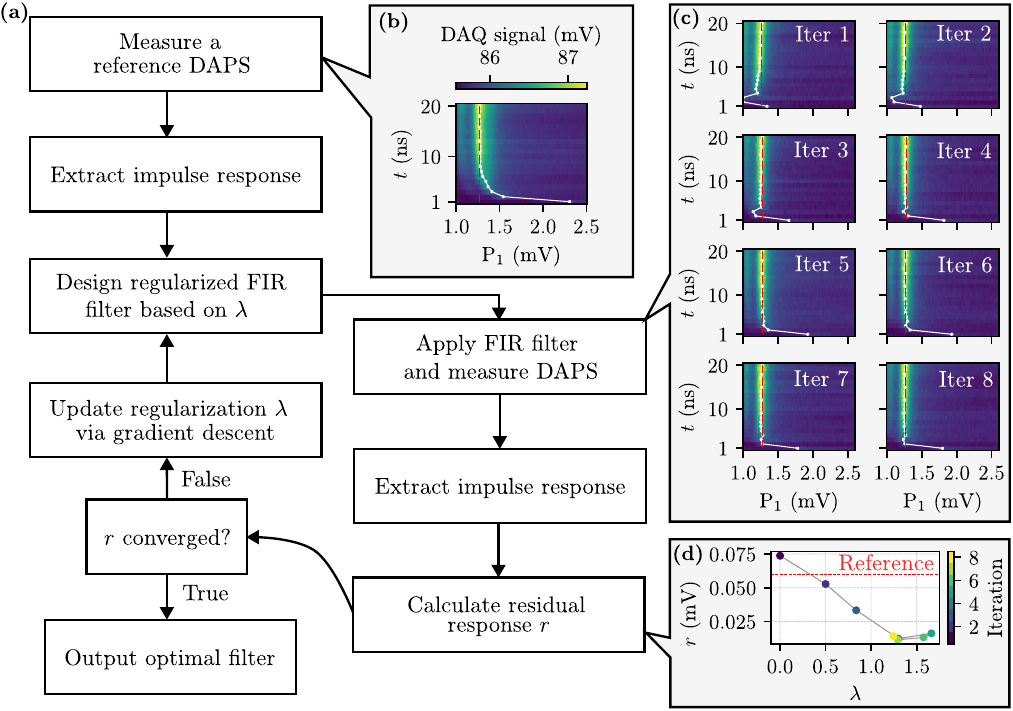}}
\caption{
(a) Flowchart of the filter optimization algorithm.
(b) Reference DAPS, where $t$ is the plateau time.
The extracted DAPS peaks are indicated by white dots, and the red dashed line represents the estimated avoided-crossing location.
(c) Pre-distorted DAPS results with iteratively updated regularization parameter $\lambda$.
(d) Residual response $r$ at each iteration of $\lambda$.
The optimal value $\lambda= 1.2924$ is reached at iteration 7.
The red dashed line indicates the residual response for the reference DAPS.
}
\label{figS:S10_optimizing_filter}
\end{figure}

While the analytical prediction of the overshoot factor in Sec.~\ref{SecS:predicting_overshoot} demonstrates that the overshoot is quantitatively understood, in practice one may prefer a fully automated approach that does not require estimating $t_c$. In this section, we demonstrate a straightforward approach to mitigate the overshoot via a filter optimization algorithm.
We use a separate device with identical geometry and wiring to the one shown in Fig.~\ref{figS:S2_calibrate_B1_B2}(a), and tune up at the (1,0)-(0,1) charge transition under gates (P$_2$, P$_1$).
We perform the algorithm shown in Fig.~\ref{figS:S10_optimizing_filter}(a) to calibrate the filter for the P$_1$ gate.
The algorithm starts by taking a DAPS measurement without filtering as a reference.
We then extract the impulse response and use it to design all filters.
We use an iterative loop to optimize the filter regularization parameter $\lambda$, as defined in Eq.~(\ref{eqS:inverse_filter_regu}).
We start with $\lambda_1=0$, design a FIR filter using Eq.~(\ref{eqS:inverse_filter_regu}), and measure a pre-distorted DAPS.
We then extract the DAPS peaks and calculate the residual response $r$, defined as the sum of the unnormalized impulse response from the returned DAPS result excluding the first nanosecond. Because the step response is not normalized during this calculation, $r$ has units of millivolts and directly quantifies the voltage-level distortion remaining after pre-distortion.
The iteration is then handled by a gradient descent algorithm with an initial boost $\lambda_\text{boost}$, such that the regularization parameter in the second iteration is $\lambda_2=\lambda_1+\lambda_\text{boost}$.
The residual responses from the first two iterations yield the gradient of $r$ along the $\lambda$-axis.
The algorithm then updates the next $\lambda$ by multiplying the current gradient of $r$ with respect to $\lambda$ by a learning rate.
The algorithm terminates when the residual response $r$ has converged.

Figure~\ref{figS:S10_optimizing_filter}(c) shows the DAPS results at each iteration, for a total of 8 iterations.
Figure~\ref{figS:S10_optimizing_filter}(d) illustrates the corresponding residual response at each iteration, where the red dashed line indicates the residual response for the reference DAPS.
We observe a significant overshoot at $\lambda=0$, as expected from the theoretical modeling.
The algorithm captures the gradient and reaches the optimal regularization parameter $\lambda=1.2924$ at iteration 7, where the residual response is $11.5~\mu$V, significantly lower than the reference case.
This result demonstrates that the overshoot mitigation reduces to a one-dimensional optimization over $\lambda$, which is straightforward to implement, time-efficient, and fully automatic.

\newpage

\end{document}